\documentclass[twocolumn,amssymb, showpacs, showkeys,amsmath, nofootinbib,tightenlines,nobibnotes,aps,prb]{revtex4}
\usepackage[dvips]{graphicx}
\usepackage[T1]{fontenc}
\usepackage{dcolumn}
\usepackage{bm}
\usepackage{natbib}
\usepackage{amssymb}
\usepackage{color}
\usepackage{dcolumn}
\newcommand{\tjs}{tunnel~junctions}

\newcommand{\figjovps}[1]{\includegraphics[width=0.45\textwidth]{#1.PS}}
\newcommand{\Rfinal}{\mathop{R\mathrm{_{final}}}}
\newcommand{\Rinitial}{\mathop{R\mathrm{_{initial}}}}
\newcommand{\Rhalf}{\mathop{R\mathrm{_{half}}}}
\newcommand{\Imax}{\mathop{I\mathrm{_{max}}}}

\newcommand{\Ip}{\mathop{I\mathrm{_{p}}}}
\newcommand{\Ima}[1]{\mbox{$I\mathrm{_{max}}=#1$ mA}}

\newcommand{\CIS}[1]{\mbox{$CIS=#1$\%}}
\newcommand{\shift}[1]{\mbox{$\delta=#1$\%}}

\begin{document}

\title{Electromigration in thin tunnel junctions with ferromagnetic/nonmagnetic: nanoconstrictions, local
heating, and direct and wind forces}

\author{J. Ventura}\email{joventur@fc.up.pt}
\author{J. B. Sousa} \email{jbsousa@fc.up.pt} \affiliation{IFIMUP and
Faculty of Sciences U. Porto, Rua do Campo Alegre,678, 4169-007,
Porto, Portugal}

\author{Y. Liu, Z. Zhang and P. P. Freitas}
\email{pfreitas@inesc-mn.pt} \affiliation{INESC-MN and IST, Rua
Alves Redol, 9-1, 1000-029 Lisbon, Portugal }

\begin{abstract}
Current Induced Resistance Switching (CIS) was recently observed in
thin tunnel junctions with ferromagnetic (FM) electrodes \emph{i.e}
FM/I/FM. This effect was attributed to electromigration of metallic
atoms in nanoconstrictions in the insulating barrier (I). Here we
study how the CIS effect is influenced by a thin non-magnetic (NM)
Ta layer, deposited just below the AlO$_x$ insulating barrier in
tunnel junctions of the type FM/NM/I/FM (FM=CoFe). Enhanced
resistance switching occurs with increasing maximum applied current
($\Imax$), until a plateau of constant CIS is reached for
$\Imax\sim65$ mA (CIS$\sim$60\%) and above. However, such high
electrical currents also lead to a large ($\sim$9\%) irreversible
resistance decrease, indicating barrier degradation. Anomalous
voltage-current characteristics with negative derivative were also
observed near $\pm\Imax$ and this effect is here attributed to
heating in the tunnel junction. One observes that the current
direction for which resistance switches in FM/NM/I/FM (clockwise) is
opposite to that of FM/I/FM tunnel junctions (anti-clockwise). This
effect will be discussed in terms of a competition between the
electromigration contributions due to the so called direct and wind
forces. It will be shown that the direct force is likely to dominate
electromigration in the Ta (NM) layers, while the wind contribution
likely dominates in the CoFe (FM) layers.
\end{abstract}


 \maketitle

\section{Introduction}
Tunnel junctions (TJ) consisting of two ferromagnetic (FM) layers
separated by an insulator (I) \cite{Moodera} show enormous potential
for a multiplicity of applications such as read
head,\cite{TJ_sensors2} strain,\cite{applications_strain} current,
position and speed \cite{applications_spinvalve_sensor} sensors or
even to detect magnetically tagged biological
specimens.\cite{applications_biological} However, probably the most
sought after application is high performance, low cost, non-volatile
magnetoresistive random access memories (MRAMs).\cite{TJ_MRAMS} In a
tunnel junction, the magnetization of one of the FM layers (pinned
layer) is fixed by an underlying antiferromagnetic (AFM) layer. The
magnetization of the other FM layer (free layer) reverses almost
freely when a small magnetic field is applied. Due to spin dependent
tunneling \cite{Tedrow_Review} one obtains two distinct resistance
(R) states corresponding to pinned and free layer magnetizations
parallel (low R) or antiparallel (high R). However, several
drawbacks are still of concern in actual MRAM submicron devices,
like cross-talk in the array configuration or the large power
consumption to generate the magnetic field to switch R. One then
aims to replace the magnetic field-driven magnetization reversal by
a Current Induced Magnetization Switching (CIMS)
mechanism.\cite{Slow, Berger} Such goal was recently achieved in
magnetic tunnel junctions \cite{CIMS_TJs,CIMS_TJs2} for current
densities \mbox{$j\sim10^7$ A/cm$^2$}. On the other hand, Liu
\emph{et al.} \cite{CIS_first} observed reversible resistance
changes induced by lower current densities \mbox{($j\sim10^6$
A/cm$^2$)} in thin FM/I/FM TJs. These changes, although initially
attributed to the CIMS mechanism, were later found \cite{CIS2} not
dependent on the relative orientation of the magnetizations of the
free and pinned layers. This effect was then called Current Induced
Switching (CIS) and is now attributed
\cite{CIS_Deac,CIS_JOV_IEEE_TN} to electromigration (EM) in
nanoconstrictions in the insulating barrier. The combination of the
tunnel magnetoresistive and CIS effects allows the use of a magnetic
tunnel junction as a three resistance state
device.\cite{CIS_JOV_APL_3states} Both CIS and CIMS effects seem to
coexist in thin magnetic tunnel junctions for
\mbox{$j\gtrsim10^6$~A/cm$^2$}. The reasons for the observed
dominance of one effect over the other are still unclear but likely
related to structural differences in the tunnel junctions. One notes
however, that electromigration can in fact limit the implementation
of the spin torque mechanism in actual devices and be a major reliability
issue in read head sensors.\cite{Diffusion_Reliability_SVs}

When a metal is subjected to an electrical field $\textbf{E}$, the
usual random diffusive motion of atoms is biased by the resulting
driving force $\textbf{F}$, and a net atomic flux can be observed.
This phenomena is known as electromigration \cite{EM_Sorbello_book}
and $\textbf{F}$ can be written as:
\begin{equation}
\textbf{F}=Z^*e\textbf{E},
 \label{eq_EM_Z*}
\end{equation}
where $Z^*$ is the effective valence and $e$ is the elementary
charge. The force acting on the migrating ion is usually separated
into two components, both linear in the external applied electrical
field:
\begin{equation}
\textbf{F}=\textbf{F}_{d}+\textbf{F}_{w}=(Z_{d}+Z_{w})e\textbf{E}.
 \label{eq_EM_Zdirect_Zwind}
\end{equation}
The direct force $\textbf{F}_{d}$ arises from the electrostatic
interactions between the electrical field and the so called direct
valence of the ion $Z_d$ ($>0$). The theoretical calculation of the
direct force is a challenging process but $Z_d\approx Z$ ($Z=$ion
valence) is usually assumed. The wind force $\textbf{F}_{w}$ results
from momentum exchange between the current carrying electrons and
the migrating ions and so it has the direction of the electron
current (opposite to the electrical field). The wind valence $Z_w$
is simply a convenient term to describe the wind force, arising from
the fact that $\textbf{F}_{w}$ is proportional to the current
density and, in an ohmic material, to $\textbf{E}$. The competition
between wind and direct forces is often dominated by the first,
which usually controls the sign and magnitude of the effective
valence $Z^*$ and the EM process.

Here we study how a Ta non-magnetic (NM) amorphous thin layer
deposited just below the insulating barrier influences the Current
Induced Switching. In a CIS cycle, the resistance commutes between
two states due to electromigration of ions from the electrodes into
the barrier (decreasing R) and from the barrier back into the
electrodes (increasing R).\cite{CIS_Deac} We can then define the CIS
coefficient as the relative difference between these two R-states.
Interestingly, the current direction for which R-switching occurs in
FM/NM/I/FM tunnel junctions is opposite to that of FM/I/FM tunnel
junctions.\cite{CIS_JOV_IEEE_TN} Using the intuitive ballistic model
of EM, we will show that the direct force is likely to dominate
electromigration in Ta (NM) layers, while the wind force dominates
in CoFe (FM) layers. The switching direction difference will be here
associated with the dominance of different EM forces (direct or
wind) in the two types of tunnel junctions referred.

The CIS coefficient was strongly enhanced by increasing the maximum
applied current ($\Imax$), reaching almost 60\% for \Ima{80}.
However, severe R-degradation occurs when $\Imax\gtrsim65$~mA.
Voltage-current characteristics show strong anomalous
non-linearities, here associated with heating effects. Comparing our
experimental results with voltage-current characteristics as
predicted by Simmons' model,\cite{Simmons_IV} we estimate that the
temperature inside the tunnel junction reaches $\sim$600 K for
\Ima{80}. Numerical results from a model of heat generation in
tunnel junctions suggest that such high temperatures can only occur
if local current densities much larger than $j=I/A$ ($I$ the
electrical current and $A$ the total tunnel junction area) exist
within the barrier. One concludes that these hot-spots concentrate
most of the current flowing through the tunnel junction stack and
are likely the reason for the occurrence of EM in the studied \tjs.

\section{Electromigration}
\begin{figure}
\begin{center}
\figjovps{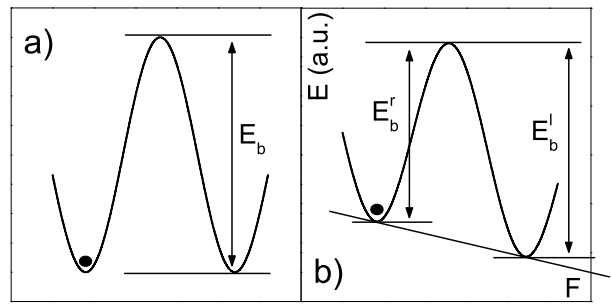}
\end{center}
  \caption{Energy barrier for atomic diffusion, a) without
  and b) with an applied electrical field. Notice how the
  direction for diffusion becomes biased by the driving force $F$:
  the energy barrier for migration to the right ($E_b^r$)
  is smaller than that for migration to the left ($E_b^l$).
  }\label{fig:Energy_barrier}
\end{figure}

For atomic diffusion to occur, an atom needs to surmount the energy
barrier $E_b$ separating neighboring equilibrium lattice sites (Fig.
\ref{fig:Energy_barrier}a). When an electric current flows through a
metal this usual, thermally-activated, random motion of atoms is
biased by the electrical field (Fig. \ref{fig:Energy_barrier}b),
resulting in a net atomic flow. This phenomena is know as
electromigration \cite{EM_Sorbello_book} and is currently the major
cause of failure of interconnects in integrated
circuits.\cite{EM_Reliability_Cu_Interconnects} Studies of EM in
interconnects are performed under severe conditions, such as high
electrical current densities \mbox{($\sim10^7$ A/cm$^2$)} and
temperature \mbox{($\sim500$--$700$ K)} and show that EM can occur
through different diffusion paths, such as grain boundary and
interfaces, as in Al \cite{EM_path_Al} and Cu \cite{EM_path_Cu}
interconnects, respectively. The relative importance of the
different diffusion paths varies with the material properties, such
as grain size and orientation, interface bonding and structure.

Electromigration is also a concern in magnetic nanostructures,
namely spin valves and tunnel
junctions.\cite{Diffusion_Reliability_SVs} During device operation,
local structural inhomogeneities can lead to large current density,
and thus to electromigration. This is of particular importance in
tunnel junctions where the resistance depends exponentially on the
barrier thickness and where localized nanoconstrictions can
concentrate most of the current. Such high current densities can
also produce intense heating leading to enhanced
electromigration.\cite{Diffusion_Reliability_SVs} Discrete
electromigration events were observed in metallic nanobridges (for
\mbox{$j\sim10^8$ A/cm$^2$}).\cite{EM_nanobridges} \emph{Reversible}
EM was recently observed in Ni nanoconstrictions
(\mbox{$j\sim10^{13}$ A/cm$^2$}) \cite{EM_nanostructures} and thin
tunnel junctions (\mbox{$j\sim10^6$
A/cm$^2$}).\cite{CIS_first,CIS_Deac} Electromigration in these
nanostructures can lead to both an increase and a decrease of the
electrical resistance, depending on the sense of the applied
electrical current, and thus on the sense of EM-driven atomic
motion.

The ballistic model of electromigration presents the most intuitive
picture of the underlying physics of EM. The wind force is
calculated assuming that all the momentum lost by the scattered
electrons is transferred to the migrating ion.\cite{EM_nanobridges}
In the free electron approximation the wind valence
becomes:\cite{EM_Sorbello_book}
\begin{equation}
Z_w=-nl\sigma_{tr},
 \label{eq_ZBallistic}
\end{equation}
where $n$ is the electron density, $l$ is the electron mean free
path and $\sigma_{tr}$ is the electron transport cross section for
scattering by the ion. Using, e.g. known values for Fe
(\mbox{$n\sim10^{-1}$ \AA$^{-3}$,} \mbox{$l\sim50$ \AA},
\mbox{$\sigma_{tr}\sim3$
\AA$^{2}$}),\cite{Ashcroft,EM_Theor_wind_force_Z} one finds
$Z_w\sim-15$ ($|Z_w|\gg Z\approx 2$). Such estimative confirms that
the wind force usually dominates electromigration. More elaborated
EM models such as the pseudopotential method give lower $Z_w$
values, by as much as 70\%.\cite{EM_Sorbello_book} However, because
of its simplicity, we will use the ballistic model to qualitatively
explain our results.

Sorbello \cite{EM_Sorbello_Mesoscopic} first studied
electromigration forces in mesoscopic systems. In particular he
considered electromigration near a point contact, modeled as a
circular aperture of radius $a$ between two metallic layers of
electrical resistivity $\rho$. He found that the direct force is
then greatly enhanced near such constriction. An estimate on the
relative magnitude of the wind and direct forces
gives:\cite{EM_Sorbello_Mesoscopic,EM_nanobridges}
\begin{equation}
\frac{F_w}{F_d}\propto -\frac{a\sigma_{tr}}{Z_d},
 \label{eq_FdFw}
\end{equation}
which evidences the important role played by the constriction
geometry: the smaller the constriction radius, the larger will be
the direct force compared to the wind force.

\section{Experimental details}
In this work we used a series of ion beam deposited tunnel
junctions, with a non-magnetic Ta layer inserted just below the
insulating AlO$_x$ barrier. The complete structure of the tunnel
junctions studied was glass/bottom lead/Ta (90 \AA)/NiFe (50
\AA)/MnIr (90 \AA)/CoFe (40 \AA)/Ta (20 \AA)/AlO$_x$ (3 \AA + 4
\AA)/CoFe (30 \AA)/NiFe (40 \AA)/Ta (30 \AA)/TiW(N) (150 \AA)/top
lead. The chosen structure is similar to that of magnetic tunnel
junctions grown for actual applications except for the additional Ta
layer, thus making a comparison between the FM/I/FM and FM/NM/I/FM
systems easier. Previous Transmission Electron Microscopy images
obtained in similar samples show no significant microstructural
changes induced by a Ta layer deposited below the
barrier.\cite{Ta_Al_oxide} The AlO$_x$ barrier was formed by
two-step deposition and natural oxidation processes (50 mTorr, 3
min, 100 mTorr, 20 min).\cite{CIS_first} NiFe, CoFe, MnIr and TiW(N)
stand for Ni$_{80}$Fe$_{20}$, Co$_{80}$Fe$_{20}$ and
Mn$_{78}$Ir$_{22}$, Ti$_{10}$W$_{90}$(N). The bottom and top leads
are made of Al 98.5\% Si 1\% Cu 0.5\%, and are \mbox{600
\AA}~(\mbox{26 $\mu$m}) and \mbox{3000 \AA}~(\mbox{10 $\mu$m}) thick
(wide) respectively. The junctions were patterned to a rectangular
shape with area \mbox{$A=4\times1$ $\mu$m$^{2}$} by a self-aligned
microfabrication process.

The electrical resistance, magnetoresistance and current induced
switching were measured with a four-point d.c. method, with a
current stable to 1:10$^{6}$ and using an automatic control and data
acquisition system.

CIS cycles were performed using the pulsed current method
\cite{CIS2} allowing us to measure the \emph{remnant} resistance of
the tunnel junction after each current pulse. Current pulses ($\Ip$)
of 1 s duration and 5 s repetition period are applied to the
junction, starting with increasing negative pulses from
\mbox{$\Ip=0$} (where we define the resistance as $\Rinitial$), in
$\Delta \Ip= 5$ mA steps up to a maximum $+\Imax$, dependent on
cycle in the 10--80 mA range. One then decreases the current pulses
(always with the same $\Delta I_p$), following the reverse trend
through zero current pulse ($\Rhalf$) down to negative $-\Imax$, and
then again to zero ($\Rfinal$), closing the CIS hysteretic cycle,
$R=R(I_{p})$. Positive current is here defined as flowing from the
bottom to the top lead.

The junction remnant resistance is measured in the \mbox{5
s}-waiting periods between consecutive current pulses, using a low
current of 0.1 mA, providing a $R(\Ip)$ curve for each cycle. This
low current method allows us to systematically discard non-linear
I(V) contributions to the resistance. However, the voltage across
the TJ is also measured while applying the current pulse $\Ip$,
enabling us to obtain the (non-linear) V($\Ip$) characteristic for
each CIS cycle.

Using the definitions above, one can define the CIS coefficient
as:
\begin{equation}\label{Defenição de CIS}
    CIS =\frac{\Rinitial-\Rhalf}{(\Rinitial+\Rhalf)/2}.
\end{equation}
We also define the resistance shift ($\delta$) in each cycle:
\begin{equation}\label{Defenição de delta}
    \delta
    =\frac{\Rfinal-\Rinitial}{(\Rinitial+\Rfinal)/2}.
\end{equation}

\section{Experimental results}
The studied tunnel junction had an initial electrical resistance
\mbox{$R=57.8$ $\Omega$} and a resistance area product
\mbox{$R\times A=230$ $\Omega\mu$m$^2$}. No magnetoresistance was
observed in our tunnel junctions, due to the loss of interfacial
polarization (20 \AA~Ta layer deposited just below the barrier). In
fact, the tunnel magnetoresistance of a TJ is known to exponentially
decrease with the thickness of a non-magnetic layer inserted just
below the insulating barrier\cite{TMRdecayCuinserted,Ta_Al_oxide}
and TMR then goes rapidly to zero within the first monolayers of the
non-magnetic material.

\begin{figure}
\begin{center}
\figjovps{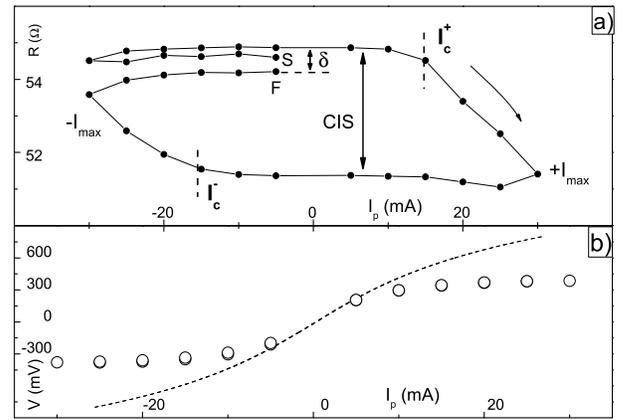}
\end{center}
  \caption{a) Current Induced Switching cycle for $\Imax=30$ mA, starting at point S and finishing at F.
  After each current pulse $\Ip$, the electrical resistance of
  the tunnel junction is measured under a low bias current, enabling us to obtain the depicted R($\Ip$) cycle.
  Effective switching occurs between $I_c^+$ and $+\Imax$, and resistance
  recovery between $I_c^-$ and $-\Imax$. b)
  Corresponding experimental (hollow circles)
  and simulated (dashed line) V($\Ip$) characteristic.
  While applying the current pulse $\Ip$, the voltage across the
  tunnel junction is measured and a V($\Ip$) characteristic obtained.}\label{fig:CIS(30)}
\end{figure}

We measured CIS cycles with increasing $\Imax$, starting with a
cycle up to \Ima{30} (Fig. \ref{fig:CIS(30)}a; cycle starting at
point S) giving \CIS{9.2} and \shift{-3.5}. No resistance switching
was observed under the initial negative current pulses
($I_p=0\rightarrow-\Imax$). However, upon reversing the current one
observes that for \mbox{$\Ip\gtrsim15$ mA} (where we define the
positive critical current $I_c^+$; see Fig. \ref{fig:CIS(30)}a) the
resistance starts to decrease, a trend which becomes increasingly
enhanced (switching) with $\Ip$, up to \Ima{30}. This switching is
associated with electromigration of metallic ions from the
electrodes into the barrier,\cite{CIS_Deac,CIS_JOV_IEEE_TN}
decreasing the effective barrier thickness and consequently the
junction resistance. The previous absence of R-switching under
negative current pulses indicates an electromigration asymmetry with
respect to the electrode/oxide interfaces, i.e. only ions from one
such interface are actively participating in electromigration.
Physically such asymmetry arises not only from the different
materials deposited just below (Ta) and above (CoFe) the insulating
barrier, but also from the deposition and oxidation processes during
tunnel junction fabrication. In particular the top electrode is
deposited over an oxidized \emph{smooth} surface, while a much more
irregular bottom electrode/oxide interface is experimentally
observed.\cite{Zhang-HRTEM} Since the migration of ions into and out
of the barrier should occur preferentially in nanoconstrictions
(higher electrical fields), one concludes that such ions likely
belong to the Ta bottom electrode. The current density and
electrical field at R-switching can be estimated as
\mbox{$j_c\sim0.375\times10^6$ A/cm$^2$} and \mbox{$E_c\sim 3$
MV/cm}, respectively.

Returning to Fig. \ref{fig:CIS(30)}a, the subsequent decrease of
$\Ip$ from $+\Imax$ to zero hardly affects the low resistance state.
However, for \mbox{$\Ip\leq-15$ mA} (where we define the negative
critical current $I_c^-$), the resistance gradually increases until
$\Ip=-\Imax$, recovering a significant fraction of the previous
R-switching near $+\Imax$. This indicates that, under a reversed
electrical field, many ions return to their initial sites. The
subsequent change of $\Ip$ from $-\Imax$ to zero (to close the CIS
cycle at point F) produces no significant change in resistance.
However, the final resistance mismatch ($\Rfinal<\Rinitial$;
\shift{-3.5}) indicates some irreversible effects in this CIS cycle
(\Ima{30}), associated with barrier degradation.

The voltage across the junction was also measured for each applied
current pulse ($\Ip$), providing the V($\Ip$) characteristic
depicted in Fig. \ref{fig:CIS(30)}b (hollow circles). If one uses
Simmons' model \cite{Simmons_IV} to fit this curve with adequate
thin TJ barrier parameters \cite{CIS_Deac} (barrier thickness
\mbox{$t=9$ \AA}, barrier height \mbox{$\phi=1$ eV}), the quality of
the fit is poor (dashed line in Fig. \ref{fig:CIS(30)}b), with large
discrepancies near $\pm\Imax$. Also, the use of the Brinkman model
for asymmetric tunnel junctions \cite{Brikman_IV} does not yield
good fits. Such discrepancies near $\pm\Imax$ are related to
localized heating inside the tunnel junction, as discussed below.

\begin{figure}
\begin{center}
\figjovps{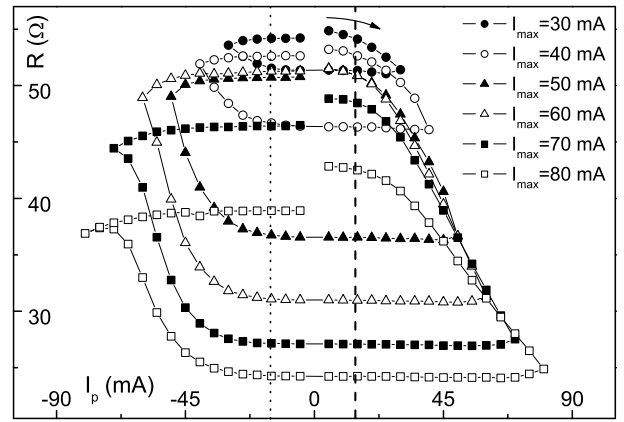}
\end{center}
  \caption{Selected CIS cycles performed with $\Imax$ up to 80 mA. Notice the enhanced
  R-switching occurring under increasing $\Imax$.}\label{fig:CIS(Imax)}
\end{figure}

We then performed CIS cycles with increasing $\Imax$, from 30 to 80
mA, in $\Delta \Imax=5$ mA steps as shown for representative cycles
in Fig. \ref{fig:CIS(Imax)}. Notice the enhanced R-switching and
R-recovering stages (versus $\Ip$), occurring from \mbox{$I_c^+$} to
$\Imax$ and from \mbox{$I_c^-$} to \mbox{$-\Imax$} respectively.
From these data one can obtain the CIS and $\delta$-shift in each
cycle, obtaining the corresponding dependence on $\Imax$ as depicted
in Fig. \ref{fig:CIS(300,perc)}. The CIS coefficient rises with
$\Imax$ until \mbox{$\sim$65 mA} (\CIS{57.4}), saturating for higher
current pulses. On the other hand, $\delta$ remains fairly small
below \mbox{$\Imax\sim60$ mA} \mbox{(-0.4\%)}, but increases rapidly
for higher $\Imax$ (\mbox{$\delta=-9.6\%$} for \Ima{80}). The CIS
increase with $\Imax$ indicates that electromigrated ions are
further pushed into the barrier (further lowering R) or/and more
ions participate in the EM processes. Ultimately irreversible damage
occurs in the barrier, as reflected in the $\delta$-shift
enhancement for \mbox{$\Imax>60$ mA} (Fig. \ref{fig:CIS(300,perc)}).

\begin{figure}
\begin{center}
\figjovps{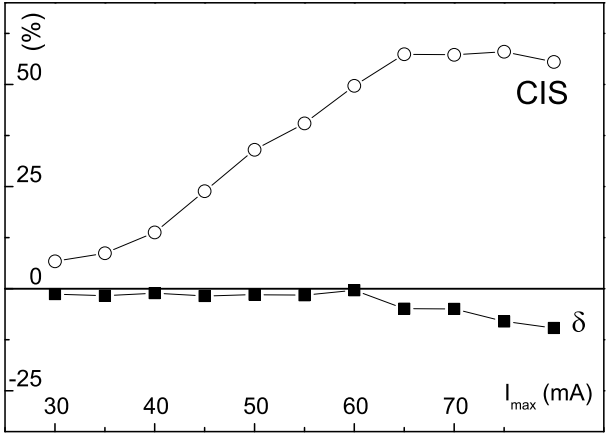}
\end{center}
  \caption{Current Induced Switching coefficient and $\delta$-shift
  as a function of maximum applied current. Large
  $\delta$-shift values occur for \mbox{$\Imax>60$ mA}, indicating progressive barrier
  degradation.}\label{fig:CIS(300,perc)}
\end{figure}

\section{Discussion}
\begin{figure}
\begin{center}
\figjovps{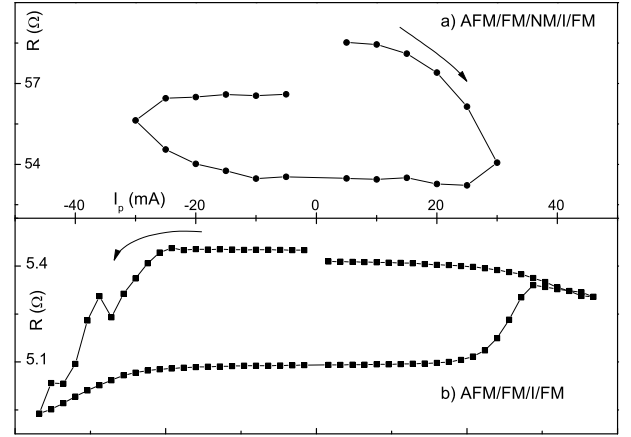}\end{center}
  \caption{Resistance switching directions for a) AFM/FM/NM/I/FM
  (MnIr/CoFe/Ta/AlO$_x$/CoFe/NiFe) and b) AFM/FM/I/FM
  (MnIr/CoFe/AlO$_x$/CoFe/NiFe) \cite{CIS_JOV_IEEE_TN} tunnel
junctions.}\label{Sentido_Inversao}
\end{figure}

The observed resistance switching (R decrease) occurs only for
\emph{positive} current pulses in the here studied FM/NM/I/FM tunnel
junctions (R-recovery occurs under negative $\Ip$; see Figs.
\ref{fig:CIS(30)}a, \ref{fig:CIS(Imax)} and \ref{Sentido_Inversao}),
whereas in the previously studied FM/I/FM \cite{CIS_JOV_IEEE_TN}
tunnel junctions switching (recovery) occurs under \emph{negative}
(positive) currents (Fig. \ref{Sentido_Inversao}b; $A=2\times1$ $\mu
m^2$). To explain such different behavior one will compare
electromigration direct and wind forces in Ta (NM) and CoFe (FM)
layers. Using eq. \ref{eq_ZBallistic} we obtain:
\begin{equation}
\frac{Z_w(Ta)}{Z_w(CoFe)}=\frac{\rho(CoFe)v_F(Ta)}{\rho(Ta)v_F(CoFe)}\frac{\sigma_{tr}(Ta)}{\sigma_{tr}(CoFe)}
 \label{eq_ZBallistic-ratio}
\end{equation}
where $v_F$ is the Fermi velocity. Inserting the parameters given in
Table \ref{Table_par_Zw}
\cite{Ashcroft,EM_Theor_wind_force_Z,FermiVel_Ta,FreitasMnIr} one
obtains $Z_w(Ta)\sim0.07 Z_w(CoFe)$. The wind force is then much
larger in CoFe than in Ta layers and likely dominates
electromigration in the CoFe layers. On the contrary, because Ta is
in an amorphous state (notice its high electrical resistivity in
Table \ref{Table_par_Zw}), one expects the small electron mean free
path to prevent large momentum gains by electrons between
consecutive collisions. Using the value estimated previously for
$Z_w(Fe)$, one finds \mbox{$Z_w(Ta)\sim-1.4$} \mbox{($\approx
Z_d(Ta)$)}. Remembering that the magnitude of the direct force is
enhanced relatively to the wind force in nanoconstrictions (eq.
\ref{eq_FdFw}; see also below) and that the ballistic model
overestimates $Z_w$, one expects the direct force to dominate in Ta.
Thus, the likely cause for the observed difference in the
R-switching directions is related to the dominance of different
electromigration forces in Ta and CoFe. Confirming this conclusion,
tunnel junctions with Ta layers deposited just below and just above
the insulating barrier (FM/NM/I/NM/FM; not shown) display the same
current switching direction as those with only one Ta layer below
the insulating barrier (FM/NM/I/FM). On the other hand, when a
single NM Ta layer is deposited just above the barrier (FM/I/NM/FM),
the R-switching direction is that of FM/I/FM tunnel junctions.

\begin{table}
\begin{ruledtabular}
\begin{tabular}{cccc}
  &$\rho$ ($\mu\Omega cm$) \cite{FreitasMnIr}&$\sigma_{tr}$
(\AA$^2$) \cite{EM_Theor_wind_force_Z}&$v_F$ (cm/s)
\cite{Ashcroft,FermiVel_Ta}\\
  \hline CoFe&17.1&$\sim$3&$\sim$2\\ Ta&154.0&$\sim$6&0.67\\
\end{tabular}
\caption{Electrical resistivity, electron transport cross section
for scattering and Fermi velocity used to estimate
$Z_w(Ta)/Z_w(CoFe)$.}\label{Table_par_Zw}
\end{ruledtabular}
\end{table}

\begin{figure}
\begin{center}
\figjovps{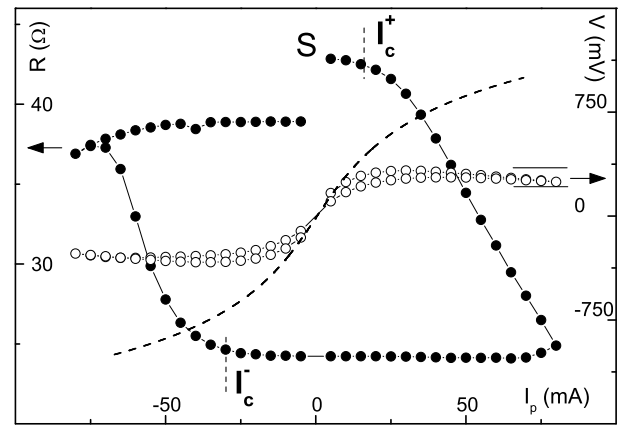}
\end{center}
  \caption{CIS cycle and corresponding V($\Ip$) characteristic
  for \Ima{80}. Notice the decrease of $|V|$ near $\pm\Imax$. The dashed line
  depicts a V(I) curve calculated using Simmons' model.}\label{fig:CIS(IV(Imax=80))}
\end{figure}

Figure \ref{fig:CIS(IV(Imax=80))} (left scale) shows the CIS
R($\Ip$)-cycle obtained at room temperature, with \Ima{80}
(\CIS{55.5}; \shift{-9.6}). Notice the R($\Ip$)-switching from
\mbox{$I_c^+=15$ mA} to \Ima{80} and resistance recovery from
\mbox{$I_c^-=-35$ mA} to \mbox{-\Ima{-80}}. The V($\Ip$)
characteristic is also displayed (hollow circles; right scale),
showing an anomalous plateau with a slight dV/d$\Ip$ negative slope
for \mbox{|$\Ip|\gtrsim30$ mA}. This effect cannot be explained by
tunnel transport theories and is here related to heating inside the
tunnel junction. Using our temperature dependent
R-data,\cite{CIS_JOV_to_be_pub_R(T)_NM} the temperature inside the
tunnel junction is estimated as $\sim$600 K. Such high temperatures
have also been observed in similar measurements performed in FM/I/FM
tunnel junctions.\cite{CIS_JOV_IEEE_TN}

Heat generation in tunnel junctions arises from two
processes:\cite{Hot_spots} usual Joule heating in the metallic
layers and inelastic electron scattering upon ballistic tunneling.
The steady-state heat equation can then be written
as:\cite{Hot_spots}
\begin{equation}
-K\frac{\partial^2T }{\partial^2x}=\rho
j^2+\frac{jV}{l_{in}}e^{-x/l_{in}}
 \label{eq_Heat}
\end{equation}
where $K$ is the heat conductivity, $T$ is the temperature, $x$ is
the stack position, $j=V/(RA)$ is the current density, $V$ is the
bias voltage and $l_{in}$ is the inelastic scattering electron mean
free path. We obtained numerical results assuming that the current
density is constant throughout the junction stack. The temperature
at the bottom and top of the tunnel junction stack is assumed fixed
at \mbox{300 K}.

\begin{figure}
\begin{center}
\figjovps{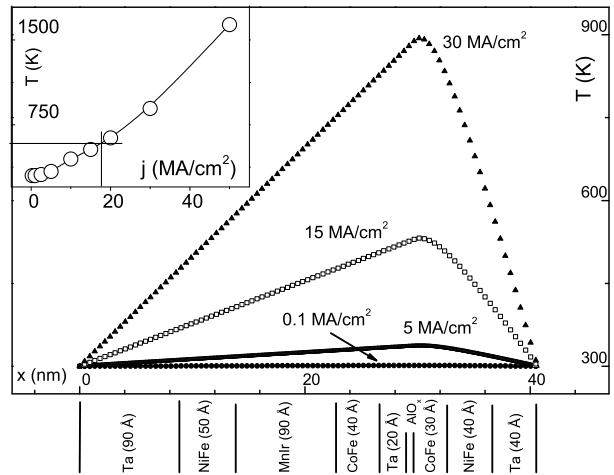}
\end{center}
  \caption{Simulation of heating processes inside the studied tunnel junction, under different
  electrical current densities (MA/cm$^2$).
  Inset: temperature increase as a function of current density passing through the junction.
  The lines show the current density needed for the temperature
  inside the junction to reach 600 K (1 MA$=10^6$ A).}\label{Fig_NumResults}
\end{figure}

Our numerical results (Fig. \ref{Fig_NumResults}) indicate that
large heating can occur near the insulating barrier for high current
densities. However, the temperature increase expected from the
uniform case, \mbox{$j_c=I_c/A\sim0.375\times10^6$ A/cm$^2$} is
negligible (\mbox{$\sim1$ K}; inset of Fig. \ref{Fig_NumResults}),
and to reach $600$ K one needs \mbox{$j_{est}\sim16\times10^6$
A/cm$^2$}. This corresponds to an effective area through which
current flows \mbox{$A_{eff}=I_c/j_{est}\approx0.1$ $\mu$m$^2$},
i.e., about 2.5\% of the total tunnel junction area. These results
then suggest that $j_c$ is only an \emph{average} value and that
nanoconstrictions where the insulating barrier is thinner
concentrate most of the current flowing through the junction. Such
hot-spots have been observed in similar TJs by atomic force
microscopy.\cite{Hot_spots}

\begin{figure}
\begin{center}
\figjovps{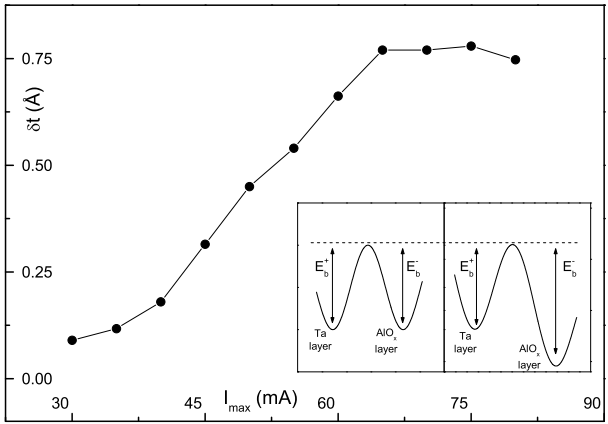}
\end{center}
  \caption{Dependence of the effective barrier thickness decrease ($\delta t$) on the maximum applied
  current pulse, as obtained from the
  $CIS(\Imax)$ curve (Fig. \ref{fig:CIS(300,perc)}) and eq. \ref{eq_DR/R(t)} (for $\phi=1$ eV). Inset: energy barrier
  for migration from Ta into the barrier ($E_b^+$) and vice-versa ($E_b^-$), in the first (left) and last (right) CIS cycles.}
  \label{Fig_Variacao_de_t_comImax}
\end{figure}

One can now understand the observed electromigration driven
resistance changes in thin FM/NM/I/FM tunnel junctions with NM=Ta
(amorphous; Fig. \ref{Sentido_Inversao}a). Under increasing positive
current pulses (directed from the bottom to the top lead), the
dominating EM direct force induced by the electrical field pushes Ta
atoms into the barrier, a process thermally assisted by heating
generated by the high current densities flowing in
nanoconstrictions. This rises the probability that an atom surmounts
the energy barrier for migration $E_b$ (see Fig.
\ref{fig:Energy_barrier}), greatly enhancing atomic mobility. One
notices that even a small barrier weakening (due to such migration)
would considerably lower the tunnel resistance due to its
exponential dependence on barrier thickness.\cite{Simmons_IV} Using
the Simmons' model we can calculate the resistance variation due to
a small barrier thickness reduction from $t$ to $t-\delta t$
($\delta t\ll t$):
\begin{equation}
\begin{split}
&\frac{R(t)-R(t-\delta
t)}{R(t)}=\\=&\frac{\Rinitial-\Rhalf}{\Rinitial}\approx
1-\-e^{-B(\phi)\delta t}\approx B(\phi)\delta t
\end{split}
 \label{eq_DR/R(t)}
\end{equation}
where $B(\phi)=0.72\sqrt{\phi/2}$. For a CIS coefficient of
$\sim60$\% one obtains a barrier thickness decrease \mbox{$\delta
t\sim0.8$ \AA}. We can now plot the magnitude of the expected
$\delta t$ decrease as a function of the maximum applied current
$\Imax$ (Fig. \ref{Fig_Variacao_de_t_comImax}; using the
experimental $\Rinitial$ and $\Rhalf$ values), which follows the
same trend as the CIS coefficient (Fig. \ref{fig:CIS(Imax)}). In
particular, a non-linear behavior (apparently exponential, as more
clearly visible at low temperatures
\cite{CIS_JOV_to_be_pub_R(T)_NM}) is observed for \mbox{$\Imax\le60$
mA}, that is, while the $\delta$-shift is small and electromigration
is mainly reversible. In atomic diffusion processes one often has
\cite{Diffusion_Prime} \mbox{$\frac{\partial x}{\partial t'}\propto
F$} ($x$ the position and $t'$ the time). Therefore, in
electromigration \mbox{$\delta t\propto E\delta t'$}, i.e. the
barrier thickness decrease is proportional to the applied electrical
field density and to the migration time $\delta t'$. Following this
simple analysis, one has $\left(R(t)-R(t-\delta
t)\right)/R(t)\propto E$. The CIS effect then depends on how local
electrical fields behave near nanoconstrictions and on its
dependence on nanostructural atomic rearrangements.

Time dependent measurements (over \mbox{4 h}) revealed that R
remains practically constant both in its high and low state (not
shown). This indicates that under a reduced driving force, displaced
Ta ions remain trapped in deep enough local energy minima inside
lattice potential barriers ($E_b\gg k_BT$), so that thermal
fluctuations cannot return them to the electrodes. For example, in
the CIS cycle of Fig. \ref{Sentido_Inversao}a one observes that the
low resistance state persists for $\Ip$ current pulses from
$\approx+\Imax$ down to $\approx I_c^-$. However, when $\Ip<I_c^-$
the driving force gets strong enough to return displaced ions back
into their initial positions in the NM layer. However, the final
resistance does not exactly reaches its initial value, indicating
progressive barrier degradation. Such degradation should result from
metallic ions that remain in the barrier after the CIS cycle is
completed. We also notice that in the initial CIS cycle with
\Ima{30} (Fig. \ref{Sentido_Inversao}a) one has $I_c^+\approx
\left|I_c^-\right|$. This indicates that the driving force for
electromigration into and out of the insulating barrier is
approximately equal, i.e. the lattice sites where ions migrate to
are energetically similar. Furthermore, Fig. \ref{fig:CIS(Imax)}
(see dashed line) shows that \mbox{$I_c^+\approx15$ mA} throughout
all the CIS cycles performed, indicating that cycling does not alter
the EM force inducing atomic migration from Ta into the barrier. In
other words, the energy barrier which the Ta ions surmount when
migrating into the barrier is kept constant (inset of Fig.
\ref{Fig_Variacao_de_t_comImax}). This contrasts with
electromigration in the opposite direction, where $|I_c^-|$
increases with cycling (Fig. \ref{fig:CIS(Imax)}; see dotted line).
The force needed to return ions back has to be increased (inset of
Fig. \ref{Fig_Variacao_de_t_comImax}), indicating that Ta ions
migrating under increasingly higher current pulses are pushed
further inside the barrier, and are thus more difficult to return to
the electrode.

\section{Conclusions}
We studied the Current Induced Switching effect on low resistance (7
\AA$~$ barrier) CoFe/Ta/AlO$x$/CoFe tunnel junctions. The CIS
coefficient increased with increasing maximum applied current
pulses, reaching $\sim$60\% for \Ima{80}. Such effect is controlled
by nanostructural rearrangements at the electrodes/barrier
interfaces, due to ion electromigration (reversible and
irreversible). When high currents are applied, one observes large
irreversible resistance decreases. The V($\Ip$) characteristics
showed an anomalous behavior when $|\Imax|\gtrsim65$ mA due to
heating effects inside the tunnel junction, showing that the CIS
effect is thermally assisted. The analysis of these effects shows
that nanoconstrictions indeed concentrate most of the tunneling
current through the barrier, forming local hot-spots. One further
demonstrates that the R-switching direction is related to a
competition between the electromigration contributions due to direct
and wind forces: the direct force dominates electromigration in Ta
layers, whereas the wind contribution is dominant in CoFe.

Finally, please notice that, although the results presented here
concern a single FM/NM/I/FM tunnel junction, they are reproduced
when measuring other TJs from the same deposition batch.
Particularly, the dependence of the CIS coefficient on maximum
applied electrical current is quite similar in different tunnel
junctions. The current switching direction is always the same for
the same TJ-structure.

\acknowledgments{Work supported in part by FEDER-POCTI/0155,
POCTI/CTM/36489/2000, POCTI/CTM/45252/02 and POCTI/CTM/59318/2004
from FCT and IST-2001-37334 NEXT MRAM projects. J. Ventura is
thankful for a FCT doctoral grant (SFRH/BD/7028/2001). Z. Zhang and
Y. Liu are thankful for FCT post-doctoral grants (SFRH/BPD/1520/2000
and SFRH/BPD/9942/2002).}

\bibliography{Biblio}

\end{document}